\documentclass[12pt]{article}

\usepackage{amssymb}

\usepackage{epsfig}

\usepackage[sf,small]{titlesec}

\usepackage{graphics}

\usepackage[latin1]{inputenc}

\begin{document}
\noindent
\renewcommand{\thefootnote}{\fnsymbol{footnote}}
\thispagestyle{empty}
\begin{center} {\Large{The impact of external events on the emergence of social herding of economic sentiment}}

\vspace{0.4cm}

{{Martin Hohnisch, Dietrich Stauffer and Sabine Pittnauer}}{\footnote{
We would like to thank F. Westerhoff and anonymous referees for very valuable comments on  a previous version of the 
paper. All conceptual and technical shortcomings of the paper are those of the authors.
Address: Stauffer: Institute of Theoretical Physics, University of Cologne, 
Z\"ulpicher Str. 77, D-50923 K\"oln, Euroland; Hohnisch and Pittnauer:  
Experimental Economics Laboratory, Department of Economics, University of Bonn, Adenauerallee 24-42, D-53113 Bonn, 
Germany (e-mail: Martin.Hohnisch@uni-bonn.de and Sabine.Pittnauer@uni-bonn.de).}}


\end{center}

\vspace{0.6cm}

{\abstract{\noindent
We investigate  the impact of an exogenous environment on the emergence of social herding of economic sentiment.  
An interactions-driven dynamics of economic sentiment is modeled by an Ising model on a large (two-dimensional) square lattice. 
The individual states are called {\it{optimism}} and {\it{pessimism}}.
The exogenous environment  is modeled as a  sequence of  random  events, which might have a positive or negative influence on economic sentiment.
These exogenous events can be frequent  or rare, have a lasting impact or a non-lasting impact. 
Impact of events is inhomogeneous over the lattice, as individuals might fail to perceive particular events. 
We introduce two notions of social herding: 
{\it{permanent herding}} refers to the situation where an ordered state (i.e. a state with an overwhelming majority of optimists or 
pessimists)  persists over an infinite time horizon, while {\it{temporary herding}} refers to the situation
where ordered states appear, persist for some time and decay.
The parameter of the inter-agent interaction strength is such as to engender permanent herding without the 
influence of the environment.
To investigate  the impact of an  environment we determine whether an initially ordered state decays.  
We consider  two cases: in the first case positive and negative events have both
the same empirical frequencies and strengths, while in the second case events have 
the same empirical frequencies but different strengths.
(In the first case the environment is ``neutral''in the long term), 
In the neutral case we find temporary herding if  events are sufficiently ``strong'' and/or perceived by a sufficiently large 
proportion of agents, and our results suggest that permanent herding occurs for small values of the parameters.
In  the ``non-neutral'' we find only temporary herding.

\vspace{0.2cm}
\noindent

}}

\section{Introduction}

\renewcommand{\thefootnote}{\arabic{footnote}}

\setcounter{footnote}{0}  

Recently, there has  been renewed strong interest  among scholars of economics in the notion of {\it{consumer sentiment}} 
-- a vague concept operationalized in particular surveys as a bundle of consumer expectations and assessments 
of the economic prospects   -- on individual and aggregate economic activity \cite{DM04,L04, So98}. 
While consumer sentiment has been considered a relevant indicator by practitioners of economic policy \cite{G02},
economic modeling is primarily concerned with other, more specific types of expectations (such as income and price 
expectations for a particular point in time), and there is much less agreement among theorists in what way -- if any -- the 
concept of consumer sentiment -- or, more generally, the concept of economic sentiment -- should enter economic modeling.

The basic assumption of the present paper is that economic sentiment  is prone 
to  imitative social influence, in that, say, a consumer is more likely  to hold an optimistic (pessimistic) expectation 
about the  economic prospects if the peers do. That assumption is well substantiated:
 according to experimental social psychology,  an individual is  the more 
likely to conform to the judgment of others the less (s)he is able to form an own judgment  
in a rational and informed manner  \cite{A51, A56, Festinger}. 
Since experimental studies in the human perception of complex dynamic systems \cite{F, S95} suggest  
that a typical consumer has only a limited perception about the functioning of the economy and the political
system in which it is embedded (arguably, this limitation applies to a considerable extent even to specialists),
one is indeed led to the conclusion that the formation of economic sentiment is prone to social imitation.
Social imitation of consumer sentiment  might or 
might not result in herding of economic sentiment at the macroscopic level. The conditions for the emergence 
of sentiment herding  at the macroscopic level are investigated for a particular model in the present paper.

The general phenomenon of social herding  has attracted much interest in economic theory over the last two decades, 
particularly in the wake of disturbances on financial markets  (see \cite{B92, BHW92, K91, K93, LM99, CB00} for some seminal contributions). 
 A principle question in models of  social herding is whether or not individual behavior should be derived from the principles of economic rationality. 
The first two of the above cited papers do so. Our paper does not, as it is based on the {\it{statistical modeling approach}}{\footnote{
See \cite{F74, sch, GGS} for early formulations  of the statistical modeling approach in economics and sociology.}}
which directly applies to sentiment formation the empirical evidence of {\it{social comparison}} processes
(see \cite{A51, A56, Festinger}) rather than explaining economic sentiment formation from rationality 
principles. We believe that the statistical approach is particularly  appropriate  
for modeling the dynamics of economic sentiment because
economic sentiment, if considered  as a particular instance, or at least part, of a consumer's {\it{mental model}} \cite{N, Sh}, 
is a premise  of individual reasoning and decision-making rather than  the subject of it.   
Most directly, our present paper belongs to the recent literature  on socially-driven  economic sentiment 
formation \cite{FF, HPSS, EHS, W, WH}.

The aim of this paper is to investigate  the impact of an exogenous environment on the emergence of social herding of consumer sentiment.  
Indeed, economic traders react as much to the news coming from the broader geo-political 
environment -- whether or not these news items are objectively interpreted -- as to the behavior/advice of others. 
Consumer sentiment subject to social imitation is modeled in our paper as a  large Ising field with nearest-neighbor interactions 
on a (two-dimensional) square lattice. The individual states are called {\it{optimism}} and {\it{pessimism}}.
Social herding corresponds to the emergence of  coordination states of the Ising model,
i.e. states with predominantly optimistic or predominantly pessimistic individual entities. 
The  environment  is modeled as a a sequence of exogenous  events (external influences), stochastically fluctuating over
time.  The exogenous events can be frequent  or rare, have a lasting impact or a non-lasting impact. 
The field of events is not homogeneous, as individual actors might fail to perceive  events.
Though the environment does have an impact in existing models of financial herding, for instance
in chartists-fundamentalists models \cite{K91, LM99} as changes in the fundamental value,  or as news affecting traders \cite{MW},
our model -- due to its simple abstract structure -- is particularly suitable for analyzing the interplay of social (local) interactions and environmental 
(global) influences in a more abstract way suitable for computer simulations.
{\footnote{ In socio-economic applications of the Ising model,  an external environment has been previously considered by
\cite{GGS, GM, I}. Relatedly, there has been much interest recently in the more general issue of disentangling endogenous  
and exogenous  dynamics in complex systems \cite{DS, S}.}}

Motivated by our results, we introduce two notions of social herding in our model: 
{\it{permanent herding}}, the stronger notion, refers to an ordered state (i.e. a state with an overwhelming majority of optimists or 
pessimists) which persists over an infinite time horizon, while {\it{temporary herding}} refers 
to a state in which ordered states appear, persist for some time and decay.
The parameter of the inter-agent interaction strength in the underlying Ising field is such as to engender 
a persistent ordered phase of the infinite model (permanent herding in our terminology) {\it{without}} the 
influence of the environment. To investigate  the impact of an  environment we determine whether an initially 
ordered state decays for the  two cases that positive and negative events have both the same empirical frequencies
and strengths (i.e. the environment is ``neutral''in the long term) and that the latter property does not hold.
In the neutral case we find temporary herding if  events are sufficiently ``strong'' and/or perceived by a sufficiently large 
proportion of agents, and our results suggest that permanent herding occurs for small values of the parameters.
In  the ``non-neutral'' case we find only temporary herding.

In  the present  paper we concentrate on the interplay of endogenous and exogenous influences on economic sentiment,
neglecting its link with real economic variables. We do so  because that link  has not been investigated at the behavioral level; 
see  {\cite{EHS, W, WH}}, however, for attempts to account for that link in similar or related models).

\section{The  model}
We let the  Ising model on a two-dimensional square lattice with nearest-neighbor interactions represent 
socially-driven  collective dynamics of economic sentiment.{\footnote{We must point out
the limitations of our basic model. First, the topology of social interactions is hardly as simple as a square lattice,
yet we are not aware of empirical investigations of network structures for our particular subject of social interactions, and
network structures found for other contexts (see e.g. \cite{B}) are not necessarily
transferable to our context \cite{Sch}. Second, interactions need not be symmetrical with respect to the individuals involved,
as is the case in our model. Third, the individual states of economic sentiment should be more rich, possibly even continuous.
However, we do believe that the Ising model provides a first approximation
to the type of systems we aim to analyze.}}
The variable $x_i$  denotes the economic sentiment of  agent $i$.
Individual states $x_i=-1$ and $x_i=1$ represent the individual states of pessimism and optimism respectively.
It is well-known  that  for  interactions between agents stronger than some critical value $J_c$  (leaving aside any 
external influences)
there exist on the infinite lattice two phases  of the  sentiment field  (``coordination states''), with the 
economic actors in each of them being either predominantly pessimistic or  
predominantly optimistic \cite{D68}. These phases are stable states which emerge -- in an appropriate formal sense -- already 
in a large enough finite system \cite{KS74} (``permanent herding'' in our terminology).

The events affecting consumer sentiment (``the environment'') at a given point in time are modeled in the present 
paper as realizations of 
a random variable $B$ with the possible values   $B=b$ (``positive'' event), or $B=-b$ (``negative'' event), 
or $B=0$ (no event) in the case of a neutral environment, and $B=-2b$ ceteris paribus in the case of a biased  environment. 
We assume an agent perceives the event correctly with probability  $p$, while ignoring the event 
with probability $1-p$. Perception of an event is independent among agents.
We introduce a variable $\epsilon_i$ such that $\epsilon_i=1$ represents the situation that agent $i$ perceives the 
event and $\epsilon_i=0$ that he does not.

According to  principles of statistical modeling, the following interaction 
potential{\footnote{In physics, Eq. 1 has the interpretation of energy, and the sum of individual energy contributions is called Hamiltonian, and the model is 
the Ising model.
In social sciences, we do not have a quantity corresponding to energy, such that
Eq. 1 is merely a representation of interactions between people and events.}}  
appropriately characterizes the interaction structure of our model in a finite square lattice $\Lambda$ 
\begin{equation}
H(x)=   -\frac{J}{2}\sum_{i,j \in \Lambda: ||i-j||=1} x_i x_j -B\sum_{i \in \Lambda} \epsilon_i x_i,
\end{equation}
with periodic boundary conditions specified in our simulations. The strictly positive parameter $J$ characterizes the interaction-to-noise ratio.
The first sum accounts for local interaction between individual agents, while the second accounts for the impact
of the exogenous events.

In Monte-Carlo Statistical Physics equilibrium states are obtained from an appropriate algorithm (which can be 
interpreted as a stochastic dynamics of the system), whereby individual sites are sequentially updated
according to the probabilities proportional to $\exp(-H)$, using the prevailing configuration of next-neighbors.
In doing so, we use the following specifications of the  process  $B_t$ representing the environment: 
external events can be frequent (time scales of the Ising field and the external field are comparable) (see Figure 1, top),
lasting but rare, (e.g. the environment may change  only once in $T$ updates of all 
individual variables) (see Figure 1, top), and rare transitory (i.e. shocks; see Figure 1 bottom). Positive and negative 
events/shocks occur equally frequent (on average) in all cases.






\section{Results}
Figure 2 summarizes our simulation results on the persistence of  an initial ordered state of the consumer sentiment 
field for the case of a large system and a neutral environment. 
The curves in Figure 2  separate areas of the parameter space -- the parameters being the proportion $p$ of agents perceiving the 
event{\footnote{Clearly, for large enough systems this fraction equals the probability $p$ of an agent 
perceiving the event.}} and event ``strength'' $b$ -- for  which the initial ordered state persists over  
 4000 Monte Carlo time steps (these areas are below a curve), and in which the initial ordered states do not
survive over 4000 MCTS (these areas are above a curve).
For a fixed $b$, the proportion $p$ was diminished until for half of
the four simulated samples no change from the initial optimistic majority
to a slight majority of pessimists was observed during 4000
iterations (sweeps through the lattice). This border point then was put into Fig.2.

As 4000 is a somewhat arbitrary time scale, we also investigated stability of an initial ordered state over very long times for frequent news and $J_c/J=0.9$.  
Figure 3 depicts the dependence on $p$ of the median time at which  the ordered state is destroyed  for a fixed exemplary value  $b=1$.
It turns out that this time tends to infinity   for $p \simeq 0.16$, close to the corresponding $p \simeq 0.20$ for 4000 iterations in Figure 2.
This suggests that there is an area of the parameter space (presumably separated by a curve running slightly below
the curve in Fig. 2) for which an ordered state is stable over infinite time horizons for a very large lattice. (Small lattices do not have sharp transitions.)
In this case, the environment does not have a destructive impact  on collective consumer sentiment.

What are the properties of the model for parameters $b$ and $p$ above a curve in Fig. 2?
The inspection of figures displaying time-paths  of the proportion of optimists/pessimists
for parameter values above a curve in Fig. 2  (available from the authors on request) shows
that the typical time-path above a curve in Figure 2 is irregular:
Periods of collective pessimism emerge, persists for some time and decay, as do periods of collective optimism.
Periods of collective pessimism and optimism  change much more slowly than the
exogeneous environment, thus collective pessimism (optimism) often persist while and despite positive (negative) events 
occur.
Periods of collective pessimism and optimism  occur -- as positive and negative events -- equally frequently over time for the
neutral environment. 

These results lead us to introduce two notions of social herding in our model: 
{\it{permanent herding}}, the stronger notion, refers to an ordered state (i.e. a state with an overwhelming majority of optimists or 
pessimists) which persists over an infinite time horizon, while {\it{temporary herding}} refers 
to a state in which ordered states appear, persist for some time and decay.
(Note that the parameter of the inter-agent interaction strength in the underlying Ising field is such as to engender 
a persistent ordered phase of the infinite model  -- permanent herding in our terminology -- {\it{without}} the 
influence of the environment.)
Thus, for the neutral environment,  our results strongly suggest that  permanently stable ordered states of collective 
pessimism/optimism do not occur if events are  too strong and/or are considered by a sufficiently large proportion of 
agents. 
This is due to a ``competition'' between the social mechanism tending to produce coordination, and the disorder of the 
external environment. Given that positive and negative events have equal empirical frequencies, such that
the environment is ``neutral'' with respect to sentiment, it is quite intuitive that a widespread perception of 
external events destroys endogenous collective states:
the disorder of the environment then prevails over the tendency to herding in economic sentiment.

We also considered an  environment which is biased in favor of pessimism. 
We take the strength of negative event as $-2b$, i.e. twice as strong  as the positive event $b$. 
Analogously to Fig. 2, the curves in Figure 4  separate areas of the parameter space in which the initial 
ordered state persists for a simulation length of 4000 Monte Carlo time steps (these areas are below a curve), and in which the initial ordered states does not
survive over 4000 MCTS (these areas are above a curve).
Fig.4 shows that the $b$ or $p$ values required for this transition are now drastically smaller than for
the neutral environments in Fig.2. 
Analogously to Fig.3 we show in Fig.5 the variation of the median destruction
time, but now for this biased case of Fig.4. Roughly the data follow a straight
line on this log-log plot, suggesting a finite decay time for all finite $p$, 
going to infinity for $p \rightarrow 0$ only. We see upward deviations for
small $p$ but the larger the lattice is the smaller are these deviations.
Thus in the biased case, in contrast to the unbiased one, the initial 
order is always destroyed if we only wait long enough. 

\section{Discussion}
States of ``collective pessimism'' -- if this social phenomenon indeed occurs -- 
might be detrimental to the efficiency of allocation of economic resources. Indeed, ``explanations'' to that effect 
can often be  heard in the public discussion about the state of the economy and economic policy.
We believe that economic-sentiment-based arguments are relevant despite the lack of proper theoretical foundations, 
and the present paper is an exploratory step toward formulating  relevant models.
Our  results confirm an intuitive presumption: attention to news reduces the prevalence of 
collective economic sentiment. This result appears to suggest that our model 
might be a useful starting point, though the present paper does not cover several important issues.
In particular, the role of the graph structure of the underlying network should be investigated.
Also, we have not specified the actual ``transmission mechanism'' of economic sentiment into economic variables
necessary for a welfare analysis of the impact of collective economic sentiment.

A more general problem lies in the fact that what we called  environment is only in part exogenous, as the economy 
itself produces relevant news which is interpreted by the decision-makers --  albeit not necessarily in a correct way. 
For instance, prolonged investor pessimism might lead to a reduction of GDP, which in turn negatively affects  investor sentiment. 
Such collective expectational biases turning into real economic forces have been qualitatively described by Keynes \cite{K}, 
but are largely neglected in modern macroeconomic theory. (In contrast, recent modeling of financial markets does
incorporate expectational biases (see \cite{L95, LM99} for  seminal contributions to this research direction).
The present model does not include  a feedback from real macro-variables (``endogenous environment'') to the economic 
sentiment (see  \cite{WH} for an attempt in this direction) as we do not specify a link of economic sentiment to real variables.

\newpage

\begin{figure}[htp]
\begin{center}
\vspace{-4cm}
\includegraphics [scale=0.9]{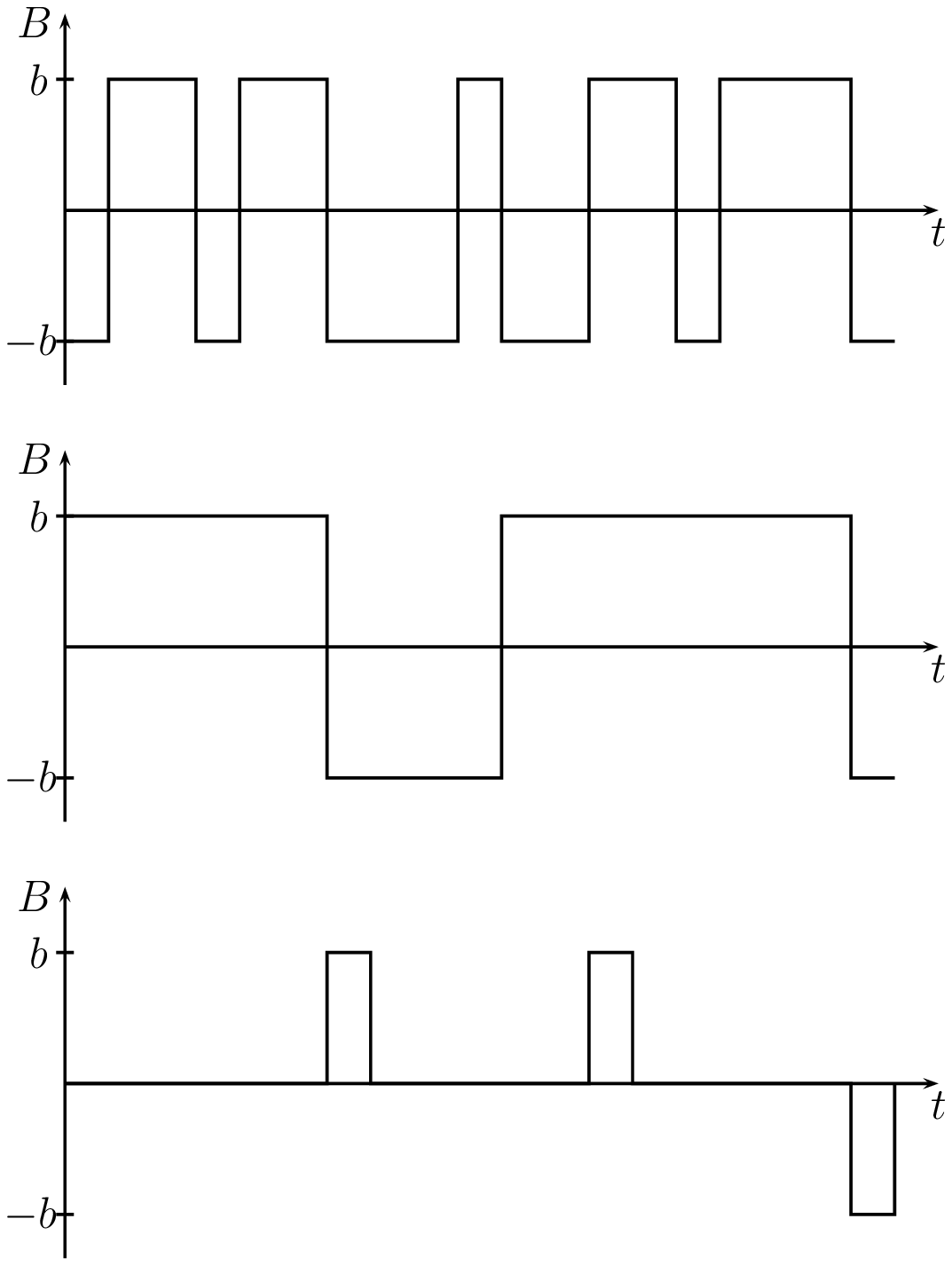}
\vspace{-5cm}
\caption{Time-patterns of the process $B_t$ representing the environment: 
time scales of the Ising field and the external field are comparable (top),
lasting but rare events (middle) and rare transitory events (shocks) (bottom). 
The proportion of positive and negative
events is equal on average for the neutral environment.}
\end{center}
\end{figure}

\newpage
\begin{figure}[htp]
\begin{center}
\includegraphics [scale=0.6,angle=-90]{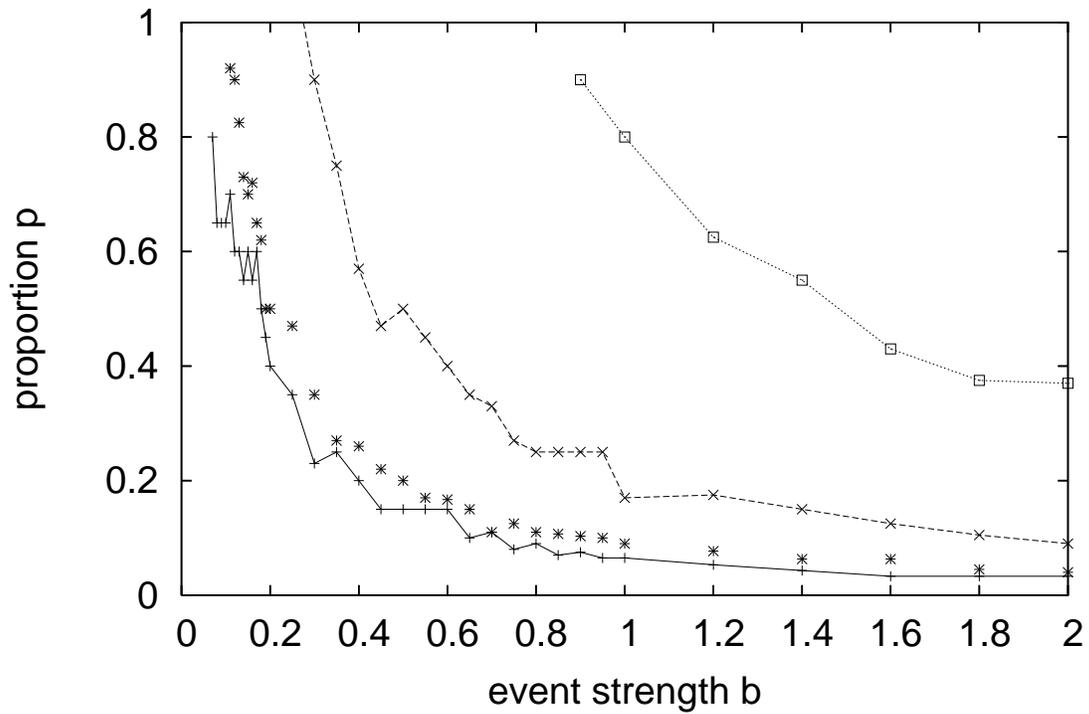}
\caption{Areas of the parameter space of our model where  ordered states persist and do not persist
for a neutral environment for 4000 sweeps through a 3001 x 3001 lattice: 
ordered states of collective sentiment exist for  vectors of the parameter values below a boundary curve, 
but do not exist for vectors of parameter values above; the curves are as follows:
($+$) frequent events, $J_c/J = 0.99$, ($\times$) frequent events, $J_c/J = 0.9$; ($*$) rare persistent events, 
$J_c/J = 0.9$, (sq.) shocks $J_c/J = 0.9$.}
\end{center}
\end{figure}

\newpage
\newpage
\begin{figure}[htp]
\begin{center}
\includegraphics [scale=0.6,angle=-90]{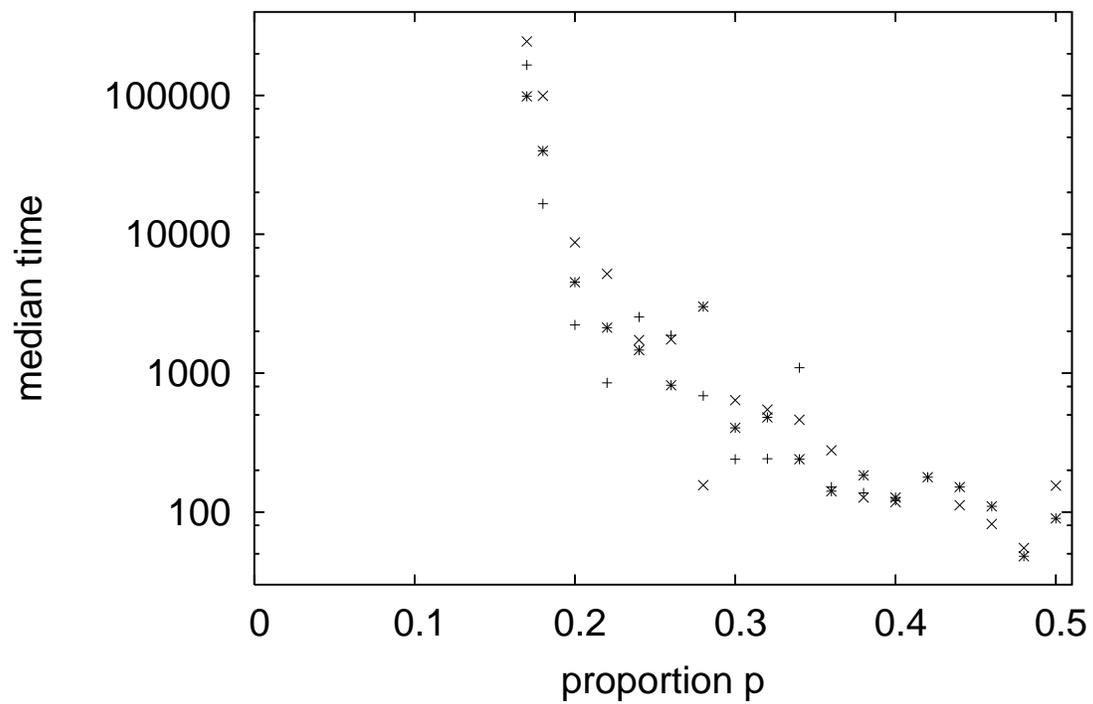}
\caption{The dependence on $p$ of the median time at which  the ordered state is destroyed  for the 
exemplary value $b=1$; frequent events, $b=1$, side of the square lattice $d= 301(+)$, 1001(x), 3001(*).}
\end{center}
\end{figure}

\newpage

\begin{figure}[htp]
\begin{center}
\includegraphics [scale=0.6,angle=-90]{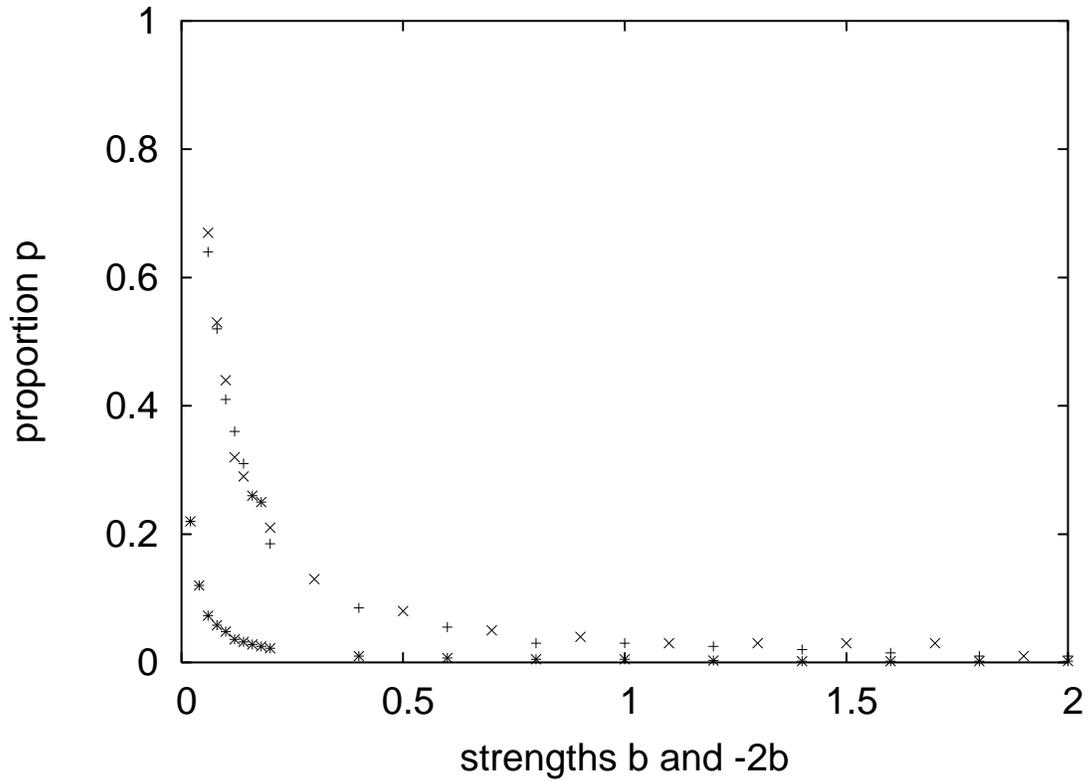}
\caption{As Figure 2, but for biased (non-neutral) environment. We made 4000 sweeps for $J_c/J$ = 0.90 (+,x) and 
0.99 (*), for frequent events only, for $1001 \times 1001$ (+) and $3001 \times 3001$
(x,*) lattices. For $1001 \times 1001$ at $J_c/J = 0.99$ the symbols would
overlap with those for the larger lattice and are thus not shown.}
\end{center}

\end{figure}

\newpage
\begin{figure}[htp]
\begin{center}
\includegraphics[scale=0.6,angle=-90]{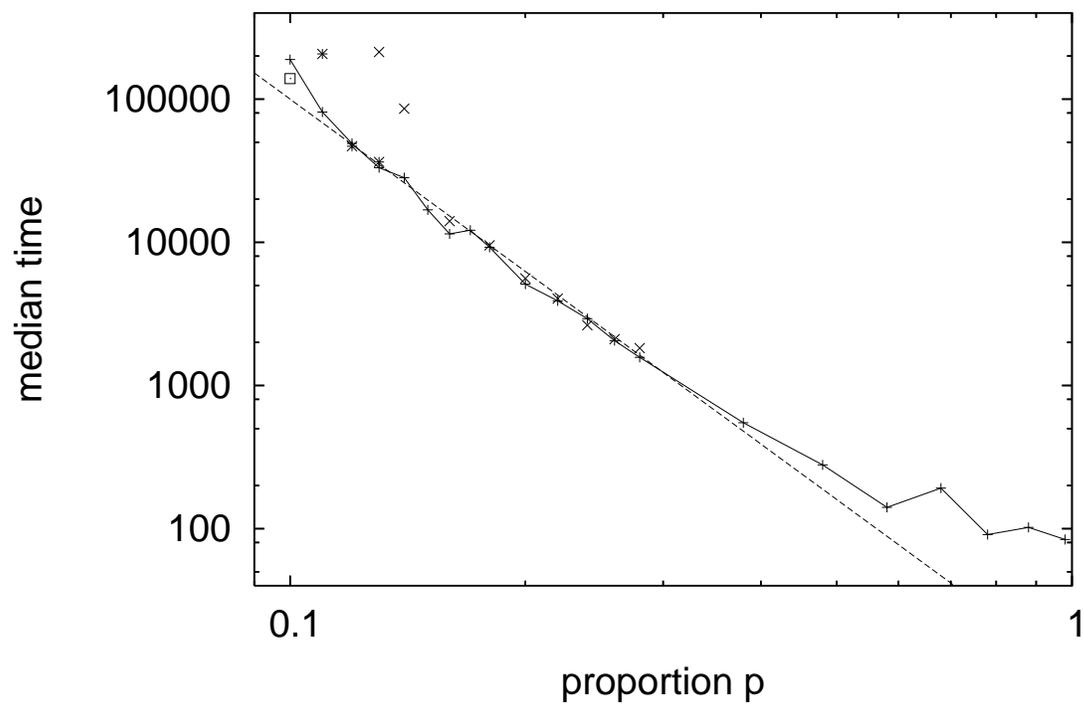}
\caption{As Figure 3, but for biased (non-neutral) environment. The side of
the square lattice $d$ is 301 (+), 1001 (x), 3001 (solid line) und 10,001
(square). The dashed line corresponds to a power law time $ = 10/p^4$. }
\end{center}
\end{figure}

\end{document}